\documentstyle[12pt]{article}
\title{A ELASTICIDADE RELATIVISTA}
\author{Ant\'{o}nio Brotas \\
{\it Instituto Superior T\'{e}cnico - Departamento de F\'{\i}sica} \\
{\it Av. Rovisco Pais, 1000 Lisboa, Portugal}\\
{\it E-mail : brotas@gtae2.ist.utl.pt}}
\date{{\it 10 de Abril de 1997}}
\begin{document}
\maketitle
\begin{abstract}

\vspace{.3cm}
The purpose of this paper is to make clear the difference 
between rigid and undeformable bodies in  Relativity. The error of 
confusing  
these two concepts has survived up to the present day treatises.

\vspace{.3cm}
We hope it will not persist in the XXI century treatises.
 
 \vspace{.3cm}
 The large majority of relativists do not know 
 the formulae for the relativistic elasticity of rigid bodies 
  presented in this paper. 
 
 \vspace{.3cm}
 The paradoxes of the rotating disk and of the 
 3-degrees of freedom of rigid bodies in Relativity  are in 
 the domain of  relativistic elasticity.

\end{abstract}

\bibliography{elasticity,rigid,indeformable}

\newpage
         
          
\begin{thebibliography}{99}
\bibitem{ } G.Herglotz."Uber die Mechanik des deformierbaren Korpers
vom Standpunkte des Relativistatstheorie".
Annalen der Physik,36,p.493,(1911)
\bibitem{ } M-Born.Annalen der Physik,30, 1,(1909)
\bibitem{ } J. Cayolla. "Sobre o movimento relativista 
de uma barra indeformavel". T\'{e}cnica, 1. Lisboa (1987) 
\bibitem{ } G.Cavalleri.Nuovo Cimento 53 B, 415 (1968)
\bibitem{ } Mc Crea,W.H. Scientific Proc. of Dublin Society, 27, 
27,(1952)
\bibitem{ } Mc Crea,W.H.and Hogarth J.E.. Proc. Camb. 
Phil. Soc.,48, 616,(1952)
\bibitem{ } A. Brotas. "Sur le probl\`{e}me du disque tournant". 
C.R. Acad. Sciences Paris. t.267 (1968)
\bibitem{ } A. Brotas.""Rigide" et "indeformable" sont-ils des 
synonimes?" 
Lettere al Nuovo Cimento. Ser. 1. p. 217. Fev (1969)
\bibitem{ } A. Brotas. "Recherches sur la Thermodynamique et la 
M\'{e}canique 
des milieux continus relativistes". 
These Fac. Sc. Paris. 
$N^{o}$ C.N.R.S. A.O. 3081 (1969) 
\bibitem{ } A. Brotas."Sobre a elasticidade relativista 
dos corpos r\'{\i}gidos". 
T\'{e}cnica 449/450. Lisboa (1978)  
\bibitem{ } A. Brotas e J.C. Fernandes. "A lei de Hooke relativista". 
T\'{e}cnica 461. Lisboa (1980)  
\bibitem{ } A. Brotas e J.C. Fernandes. "A equa\c{c}\~{a}o relativista 
do movimento de uma barra n\~{a}o r\'{\i}gida." T\'{e}cnica    . Lisboa 
(198 )
\bibitem{ } L. Bento. "Transverse Waves in Relativistic Rigid Body". 
International Journal of Theorical Physic, vol. 24, $n^{o}\ 6$, (1985)
\bibitem{ } Aldo Bressan. "Relativistic Theories of Materials". 
Springer, (1978)
\bibitem{ } A. Lichnerowicz. "Les th\'{e}ories relativistes de la 
gravitation et de l'electromagnetisme" . Masson,  Paris 1955
\end{thebibliography}

 \vspace{.4cm}  
{\bf O objectivo deste texto \'{e} impedir que nos
tratados de Relatividade do S\'{e}culo XXI se cometa o erro que tem vindo
continuadamente a ser repetido nos tratados actuais de confundir
r\'{\i}gido com indeform\'{a}vel.

\vspace{.4cm}
Os bem conhecidos paradoxos dos tr\^{e}s graus de
liberdade dos corpos r\'{\i}gidos em Relatividade e do disco a
rodar s\~{a}o problemas de elasticidade.

\vspace{.4cm}
A esmagadora maioria dos relativistas n\~{a}o conhece as leis
el\'{a}s-
ticas dos corpos r\'{\i}gidos aqui apresentadas.} 

 \vspace{.4cm}

\newpage
\section{Introdu\c{c}\~{a}o}

Na realidade h\'{a} corpos mais ou menos r\'{\i}gidos. Nos livros de
F\'{\i}sica a express\~{a}o {\it corpo r\'{\i}gido}  quando usada
isoladamente \'{e}, no entanto, entendida no sentido de {\it corpo o mais
r\'{\i}gido poss\'{\i}vel} , ou seja, {\it r\'{\i}gido limite} . Nada nos
impede em F\'{\i}sica Cl\'{a}ssica de considerar este corpo
r\'{\i}gido limite como indeform\'{a}vel, isto \'{e}, como corpo que
n\~{a}o sofre deforma\c{c}\~{o}es quaisquer que sejam as for\c{c}as
que lhe s\~{a}o aplicadas. Em F\'{\i}sica Cl\'{a}ssica r\'{\i}gido e
indeform\'{a}vel s\~{a}o assim considerados como sin\'{o}nimos,
embora r\'{\i}gido seja uma no\c{c}\~{a}o f\'{\i}sica e
indeform\'{a}vel uma no\c{c}\~{a}o geom\'{e}trica.

\vspace{.4cm}  
Nos tratados de Mec\^{a}nica Cl\'{a}ssica h\'{a} sempre dois
cap\'{\i}tulos, a Cinem\'{a}tica e a Din\^{a}mica dos corpos
r\'{\i}gidos, em que estes corpos s\~{a}o encarados como sistemas com
6 graus de liberdade. S\'{o} no seguimento destes estudos se aborda a
Elasticidade em que se consideram corpos mais ou menos r\'{\i}gidos
e, s\'{o} depois, normalmente, se estuda a Mec\^{a}nica dos Fluidos.

\vspace{.4cm}  
Nos tratados de Relatividade n\~{a}o h\'{a} cap\'{\i}tulos que
correspondam \`{a} Cine-
m\'{a}tica e \`{a} Din\^{a}mica dos corpos
r\'{\i}gidos indeform\'{a}veis da Mec\^{a}nica Cl\'{a}ssica e, em 
nenhum que eu conhe\c{c}a, h\'{a} um cap\'{\i}tulo aut\'{o}nomo sobre a
Elasticidade dos s\'{o}lidos. No entanto, em muitos, h\'{a}
cap\'{\i}tulos sobre Fluidos Relativistas.

\vspace{.4cm}  
Esta estranha situa\c{c}\~{a}o tem que ver com uma distin\c{c}\~{a}o
que tem sido igno-
rada e com um erro de designa\c{c}\~{a}o que tem
vindo a ser repetido nos tratados de Relatividade desde o
princ\'{\i}pio do s\'{e}culo.

\vspace{.4cm}  
Podemos em F\'{\i}sica Cl\'{a}ssica conceber o corpo mais r\'{\i}gido
poss\'{\i}vel, o corpo r\'{\i}gido limite, como indeform\'{a}vel.
Tal n\~{a}o \'{e}, no entanto, poss\'{\i}vel em Relatividade porque
nesse corpo-r\'{\i}gido-limite-indeform\'{a}vel as ondas de choque se
propagariam com velocidade infinita.

 \vspace{.4cm}  
{\bf \'{E} necess\'{a}rio, em consequ\^{e}ncia, fazer em Relatividade
distin\c{c}\~{a}o entre r\'{\i}gido e indeform\'{a}vel.}

\vspace{.4cm}  
Ora, o ``corpo r\'{\i}gido'' (the rigid body) referido em todos os
tratados de Relatividade \'{e}, de facto, o corpo relativista
indeform\'{a}vel que Born estudou, e muito bem, em 1909, mas
cometendo o erro de o designar por ``corpo r\'{\i}gido'' em vez de o
designar por corpo indeform\'{a}vel.

\vspace{.4cm}  
Este erro de designa\c{c}\~{a}o encobriu a necessidade de definir e
estudar o verdadeiro corpo r\'{\i}gido relativista, distinto do corpo
indeform\'{a}vel, e deu origem a uma s\'{e}rie de dificuldades e
paradoxos que ocuparam a literatura relativista  quase desde o seu 
in\'{\i}cio
e n\~{a}o foram, ainda hoje, claramente esclarecidos em nenhum tratado.

\section{Um pouco de Hist\'{o}ria}

As equa\c{c}\~{o}es de Maxwell, invariantes nas
transforma\c{c}\~{o}es de Lorentz mas n\~{a}o nas
transforma\c{c}\~{o}es de Galileu, eram equa\c{c}\~{o}es relativistas
conhecidas antes da Relatividade. O Electromagnetismo ficou, assim,
de imediato, conciliado com as novas concep\c{c}\~{o}es de espa\c{c}o
e tempo de Einstein de 1905. Para manter o Princ\'{\i}pio da
Relatividade -\it{as leis da F\'{\i}sica s\~{a}o as mesmas em todos os
referenciais de in\'{e}rcia}\rm\ - o problema passou a ser o de modificar
os outros cap\'{\i}tulos da F\'{\i}sica, invariantes nas
transforma\c{c}\~{o}es de Galileu, para ficarem invariantes nas
transforma\c{c}\~{o}es de Lorentz.
 
 \vspace{.4cm} 
Para a Mec\^{a}nica das part\'{\i}culas pontuais o problema foi
resolvido por Einstein, logo em 1905. Esta quest\~{a}o revelou-se uma
das mais fecundas da Hist\'{o}ria da F\'{\i}sica pois a sua
resolu\c{c}\~{a}o ``obrigou'' ao estabelecimento da
equipara\c{c}\~{a}o entre massa e energia.  Einstein interessou-se,
em seguida, pelo problema da Gravita\c{c}\~{a}o. \'{E} sabido como,
para o resolver, foi obrigado a ultrapassar o 
quadro do espa\c{c}o-tempo de Minkovski 
e a considerar um espa\c{c}o-tempo riemaniano.
 
 \vspace{.4cm} 
\`{A} margem ficou o problema da Elasticidade, sem d\'{u}vida muito
menos importante para o desenvolvimento da F\'{\i}sica.
  
  \vspace{.4cm} 
Um f\'{\i}sico muito capaz, G. Herglotz, debru\c{c}ou-se, no entanto,
sobre o problema da formula\c{c}\~{a}o de uma Elasticidade
Relativista e quase o resolveu, de uma s\'{o} vez, num importante
artigo com 40 p\'{a}ginas, publicado em 1911 [1], que s\'{o} tem uma
falha no final.
 
 \vspace{.4cm} 
Herglotz conhecia muito bem a Elasticidade e a Mec\^{a}nica
Anal\'{\i}tica do seu tempo, incluindo a Mec\^{a}nica Anal\'{\i}tica
dos meios cont\'{\i}nuos, e compreendeu muito bem a Relatividade. No
referido artigo desenvolveu a Elasticidade Relativista por um
m\'{e}todo variacional a partir de uma fun\c{c}\~{a}o lagrangeana ${\cal 
L}$
admitida, mas n\~{a}o precisada no in\'{\i}cio.
  
  \vspace{.4cm} 
No final, quando pretendeu aplicar a teoria desenvolvida a um caso
concreto, na falta de uma indica\c{c}\~{a}o ou crit\'{e}rio que o
orientasse, adoptou a fun\c{c}\~{a}o lagrangeana de Elasticidade
cl\'{a}ssica e linear. Esta escolha conduziu a resultados
incompat\'{\i}veis com a Relatividade, como seja a exist\^{e}ncia de 
pontos
materiais com velocidades superiores a $c$. Em consequ\^{e}ncia, o
trabalho de Herglotz foi abandonado e quase esquecido.
   
   \vspace{.4cm} 
Houve, ent\~{a}o, um recuo para 1909. Nessa data Born tinha definido
e estudado o corpo indeform\'{a}vel relativista [2]. O seu trabalho
consiste essencialmente no seguinte:
    
    \vspace{.4cm} 
Sendo $(x^i, t)$ as coordenadas espaciais e temporal de um
referencial de in\'{e}rcia $S$, considerado um ``meio cont\'{\i}nuo''
definido por um conjunto de fun\c{c}\~{o}es $x^i = x^i (X^j, t )$ com
os $X^j$ ``coordenadas fixas do meio'', \'{e} facil conhe-
cer as velocidades em $S$ dos pontos do ``meio'', em geral 
diferentes de ponto para ponto e vari\'{a}veis no tempo. Considerando 
transforma\c{c}\~{o}es de Lorentz locais, \'{e} poss\'{\i}vel definir  
a ``m\'{e}trica pr\'{o}pria do meio'' que vem
descrita por f\'{o}rmulas do tipo:

\vspace{.4cm} 
\begin{equation}
dl^{2}=\gamma_{ij}\  dX^{i} dX^{j},\ com\  \gamma_{ij}=\gamma_{ij}(X^{k}, 
t ).
\end{equation}
 
 \vspace{.4cm} 
No caso particular dos $\gamma_{ij}$ n\~{a}o dependerem de $t$, o 
``meio''
deve ser considerado indeform\'{a}vel. A f\'{o}rmula:

\begin{equation}
\left(\frac{\partial \gamma_{ij}}{\partial\  t}\right)_{X^{k}}=0
\end{equation}

define, pois, o meio ou corpo indeform\'{a}vel de Born, s\'{o} com a
diferen\c{c}a de ele lhe ter chamado r\'{\i}gido 
\footnote
{Considerados os sistemas de coordenadas : $(x^i, x^4=c\  t)$ e
$(X^i, X^4=x^4=c\ t)$, a m\'{e}trica do espa\c{c}o tempo de Minkovski
pode-se escrever: 
$ds^{2} = \sum (dx^i)^2 - (dx^4)^2 = G_{\alpha\beta}(X^k, t )
dX^{\alpha} dX^{\beta}$. 
Considerando a diferencial (em geral n\~{a}o
integr\'{a}vel) de um tempo pr\'{o}prio local $dt_{p}$ 
definido por:
$dX^{4'} = c\  dt_{p} = (-G_{i4} dX^i).(-G_{44})^{-1/2} + 
(-G44)^{1/2} dX^{4}$ 
podemos escrever: $ds^2 = \gamma_{ij} dX^i dX^j
- (dX^{4'})^2$; com $\gamma_{ij} = G_{ij}- G_{i4} G_{j4}G_{44}^{-1}$. 
Estes coeficientes $\gamma_{ij}$ d\~{a}o-nos, em Relatividade, a 
m\'{e}trica
pr\'{o}pria do meio:$dl^2 = \gamma_{ij} dX^idX^j$.}.
  
\vspace{.4cm}
\newpage

\section{Os paradoxos}
  
\vspace{.4cm}
Em F\'{\i}sica Cl\'{a}ssica o movimento de um corpo pode ser encarado
com uma sucess\~{a}o de estados est\'{a}ticos. Podendo as
posi\c{c}\~{o}es de um corpo r\'{\i}gido- (indeform\'{a}vel) num
referencial ser definidas por 6 par\^{a}metros, o seu movimento
pode ser descrito pelo movimento de um ponto no espa\c{c}o de
dimens\~{a}o 6 das configura\c{c}\~{o}es e a sua velocidade, num
sentido generalizado, por um vector deste espa\c{c}o.
 
\vspace{.4cm}
A velocidade generalizada, o estado de movimento ou, numa
descri\c{c}\~{a}o mais pormenorizada, o campo das velocidades num
dado instante dos pontos de um corpo r\'{\i}gido-indeform\'{a}vel,
s\~{a}o, assim, caracterizados por 6 par\^{a}-
metros.
  
\vspace{.4cm}
A express\~{a}o: \it{``os corpos r\'{\i}gidos t\^{e}m 6 graus de
liberdade''\rm \ tem, assim, em F\'{\i}sica Cl\'{a}ssica, um duplo 
sentido
est\'{a}tico e cinem\'{a}tico \footnote{Em F. C., descrito o 
movimento de um meio cont\'{\i}nuo por um
conjunto de f\'{o}rmulas $x^i= x^i(X^i, t)$, \'{e} facil ver se se trata
de um meio indeform\'{a}vel e, caso sim, escolher 6 par\^{a}metros
para caracterizar o campo das suas velocidades em cada instante. Em
qualquer caso, deforma-se ou n\~{a}o o meio, a sua m\'{e}trica dada
por $dl^2 = \sum dx_i^2 = g_{ij} dX^i dX^j$ \'{e} sempre euclideana. 
\'{E} a
m\'{e}trica do referencial onde \'{e} estudado o movimento e em F.C.
todos os referenciais, de in\'{e}rcia ou n\~{a}o de in\'{e}rcia,
t\^{e}m m\'{e}trica euclideana.}.
   
\vspace{.4cm}
Em Relatividade, devido \`{a} ``contra\c{c}\~{a}o de Lorentz'', o
movimento j\'{a} n\~{a}o pode ser encarado como uma sucess\~{a}o de
estados de imobilismo. Embora em Relatividade Restrita as
posi\c{c}\~{o}es im\'{o}veis de um corpo indeform\'{a}vel num referencial
de in\'{e}rcia continuem a ser definidas por 6 par\^{a}metros, o seu
movimento j\'{a} n\~{a}o pode ser descrito pelo movimento de um ponto
no espa\c{c}o de dimens\~{a}o 6 das configura\c{c}\~{o}es.
  
\vspace{.4cm}
A situa\c{c}\~{a}o \'{e} mais complexa. Para estudar os
poss\'{\i}veis movimentos de um corpo r\'{\i}gido-(indeform\'{a}vel)
\'{e} necess\'{a}rio usar a f\'{o}rmula de Born referida na 
sec\c{c}\~{a}o
anterior.
  
\vspace{.4cm}
Uma coisa sabemos, no entanto, desde o in\'{\i}cio. Tendo um corpo
r\'{\i}gido- indeform\'{a}vel um movimento poss\'{\i}vel num
referencial de in\'{e}rcia $S$, pode ter movimentos iguais em todos os
outros referenciais de in\'{e}rcia. A estes movimentos correspondem,
em $S$, movimentos que podemos obter por composi\c{c}\~{a}o do
movimento inicial com transla\c{c}\~{o}es uniformes caracterizadas
por 3 par\^{a}metros. Usando a f\'{o}rmula de Born pode-se mostrar
que estes movimentos s\~{a}o os \'{u}nicos poss\'{\i}veis.
  
  \vspace{.4cm}
\'{E} neste sentido que se diz que \it{``os corpos r\'{\i}gido
indeform\'{a}veis t\^{e}m em Relatividade 
Restrita 3 graus de liberdade''\rm\ 
\footnote{ J. Cayolla mostrou [3] que \'{e} poss\'{\i}vel, mantendo a
m\'{e}trica plana durante todo o processo, acelerar em Relatividade
Restrita um corpo limitado indeform\'{a}vel  de m\'{e}trica plana,
inicialmente im\'{o}vel num referencial de in\'{e}rcia, de modo a
ficar im\'{o}vel num outro referencial de in\'{e}rcia. Os corpos
r\'{\i}gidos-indeform\'{a}veis \it{limitados}\rm\  t\^{e}m, pois, 
um pouco mais de
liberdade do que \'{e} indicado nesta frase corrente.}.
 
 \vspace{.4cm}
Como compreender esta passagem dos 6 para os 3 graus de liberdade?

 \vspace{.4cm}
D\'{a} a impress\~{a}o que em Relatividade Restrita os corpos
r\'{\i}gido-(indefor-
m\'{a}veis) andam em calhas que limitam as suas
possibilidades de movimento.
 
 \vspace{.4cm} 
Como compreender esta limita\c{c}\~{a}o de que n\~{a}o h\'{a} sinal
em F\'{\i}sica Cl\'{a}ssica e que as experi\^{e}ncias com os
s\'{o}lidos reais de modo algum revelam?
 
 \vspace{.4cm}
Sobre este paradoxo dos tr\^{e}s graus de liberdade dos corpos
r\'{\i}gidos em Relatividade escreveram-se centenas de artigos.
 
\vspace{.4cm}
\hspace{5cm}
  x x x 

\vspace{.4cm}
Sejam $(r, \theta, z, t)$ as coordenadas espaciais cil\'{\i}ndricas e
temporal de um referencial de in\'{e}rcia $S$, cuja m\'{e}trica, em
Relatividade Restrita como em F\'{\i}sica Cl\'{a}ssica, \'{e}
euclideana. Consideremos um ``meio'' de coordenadas 
``fixas''$( R, \Theta, Z )$ 
  limitado aos intervalos 
  $( 0\leq R \leq R_{L}\ ;\ 0\leq \Theta \leq 2\pi \ ;\  0\leq Z \leq 
Z_{L} )$
cujo movimento em rela\c{c}\~{a}o a um referencial $S$ seja dado
pelas f\'{o}rmulas: $ r = R \ ;\  \theta = \Theta + \omega t \ ;\  z = Z$ 
.                              

  \vspace{.4cm} 
Os pontos deste ``meio'' descrevem c\'{\i}rculos em $S$. Em conjunto
representam um disco que, para os observadores de $S$, roda em torno do
eixo dos zz com a velocidade angular $\omega$.
 
 \vspace{.4cm}
Os c\'{a}lculos feitos com a f\'{o}rmula de Born indicam-nos que a
sua m\'{e}trica espacial pr\'{o}pria \'{e} dada por :

\vspace{.4cm}
\begin{equation}
dl^{2}=dR^{2}+\frac{R^{2}}{\sqrt{1-\beta^{2}}} \ d\Theta^{2}, 
\ com \  \beta=\frac{R\omega}{c}.
\end{equation}

\vspace{.4cm}
Sendo estes coeficientes $\gamma{ij}$ independentes do tempo, este meio
pode ser identificado com um corpo r\'{\i}gido-indeform\'{a}vel. 
A sua m\'{e}trica, por\'{e}m, n\~{a}o \'{e} euclideana.

\vspace{.4cm}
A manter-se o corpo indeform\'{a}vel n\~{a}o pode parar, porque para
parar a sua m\'{e}trica teria de passar a euclideana para coincidir
com a m\'{e}trica de $S$ e, para isso, os $\gamma_{ij}$ teriam de variar.
 
 \vspace{.4cm}
Nem pode alterar a sua velocidade de rota\c{c}\~{a}o, pois qualquer
altera\c{c}\~{a}o do movimento definido pelas f\'{o}rmulas iniciais
implicaria uma altera\c{c}\~{a}o dos  $\gamma_{ij}$ interdita aos corpos
r\'{\i}gido-(indeform\'{a}veis) de Born.
  
  \vspace{.4cm}
Este   paradoxo do disco a rodar, ou de Ehrenfest que o
assinalou em primeiro lugar, n\~{a}o \'{e} mais do que um caso
particular do anterior. 
 
 \vspace{.4cm}
\hspace{5cm}
  x x x
 
 \vspace{.4cm}
Por volta dos anos 60, numa altura em que acompanhei mais de perto
esta quest\~{a}o, constatei que, em grande maioria, os autores que
escreviam sobre o assunto procuravam abordar o problema no \^{a}mbito
da Relatividade Generalizada.
  
  \vspace{.4cm}
Procuravam, no fundo, \`{a} custa das deforma\c{c}\~{o}es do
espa\c{c}o-tempo, encontrar em Relatividade Generalizada os 6 graus
de liberdade que n\~{a}o encontravam em Relatividade Restrita.
  
  \vspace{.4cm}
O recurso \`{a} R.G. n\~{a}o facilita por\'{e}m as coisas, antes pelo
contr\'{a}rio.
   
   \vspace{.4cm}
Se considerarmos um corpo r\'{\i}gido-indeform\'{a}vel com
m\'{e}trica pr\'{o}pria pla-
na a deslocar-se com um movimento de
transla\c{c}\~{a}o uniforme numa regi\~{a}o em que o espa\c{c}o-tempo
seja plano e esse corpo chegar a uma regi\~{a}o em que o 
espa\c{c}o-tempo seja curvo, encalha na deforma\c{c}\~{a}o 
do espa\c{c}o-tempo e n\~{a}o pode prosseguir.
  
\vspace{.4cm}
Para o fazer, teria de alterar a sua m\'{e}trica pr\'{o}pria para se
adaptar \`{a} deforma\c{c}\~{a}o do espa\c{c}o-tempo, o que lhe \'{e}
interdito pela sua qualidade de ser r\'{\i}gido. 

\vspace{.4cm}
Verdade se diga que tamb\'{e}m n\~{a}o pode parar.
   
\vspace{.4cm}
E n\~{a}o nos  \'{e} poss\'{\i}vel, na sequ\^{e}ncia e no \^{a}mbito
de um estudo puramente cinem\'{a}tico como \'{e} o de Born, em que os
corpos (os ``meios'') s\~{a}o descritos pelos seus movimentos e nem
sequer s\~{a}o referidas as suas massas, utilizar as equa\c{c}\~{o}es 
de Einstein para estudar o efeito dos
corpos sobre a m\'{e}trica do espa\c{c}o-tempo.

\section{Os modelos el\'{a}sticos}

N\~{a}o h\'{a} experi\^{e}ncias de F\'{\i}sica Cl\'{a}ssica e
experi\^{e}ncias de Relatividade. As experi\^{e}ncias devem poder ser
descritas e interpretadas, tanto quanto poss\'{\i}vel por modelos
matem\'{a}ticos, no quadro da F\'{\i}sica Cl\'{a}ssica e no da
Relatividade, sendo normalmente mais aprofundadamente compreendidas
no segundo onde temos a considerar os dois n\'{\i}veis da Relatividade
Restrita e da Relatividade Generalizada (podendo, eventualmente, virem 
a ser considerados outros no futuro).
  
  \vspace{.4cm}
Podemos rodar uma moeda e descrever esta experi\^{e}ncia no quadro
da F\'{\i}sica Cl\'{a}ssica. Como admitir que esta t\~{a}o 
elementar\'{\i}ssima experi\^{e}ncia
n\~{a}o possa ser descrita no quadro da Relatividade Restrita?

 \vspace{.4cm}
 E se o for, como compreender que o deixe de ser na passagem ao limite
da ``moeda r\'{\i}gida''?
  
  \vspace{.4cm}
Podemos admitir que a descri\c{c}\~{a}o s\'{o} seja poss\'{\i}vel no
quadro da Relatividade Generalizada?
 
 \vspace{.4cm}
Se fosse o caso, a partir de uma descri\c{c}\~{a}o em Relatividade 
Generalizada  e fazendo
tender para zero a constante que figura nas equa\c{c}\~{o}es de
Einstein, teriamos uma descri\c{c}\~{a}o em Relatividade Restrita.
  
  \vspace{.4cm}
Pareceu-me claro que o problema era bem um problema de Relatividade
Restrita, e que as tentativas para o resolver em Relatividade
Generalizada eram fugas para diante. 

 \vspace{.4cm}
Um dia, em 1963, quando passeava numa praia no Recife, 
ao fazer a pequena experi\^{e}ncia de rodar uma moeda notei 
algo que nunca vi referido nos livros e foi para mim a 
chave para esclarecer o problema.
   
   \vspace{.4cm}
Uma moeda parada e a mesma moeda a rodar, s\~{a}o a mesma moeda {\it ``em
condi\c{c}\~{o}es f\'{\i}sicas diferentes'}'. Uma moeda parada e a
mesma moeda num movimento de transla\c{c}\~{a}o uniforme, s\~{a}o a
mesma moeda {\it ``em id\^{e}nticas condi\c{c}\~{o}es f\'{\i}sicas''}.
Os conjuntos de poss\'{\i}veis estados de um corpo s\'{o}lido {\it em
id\^{e}nticas condi\c{c}\~{o}es f\'{\i}sicas} t\^{e}m em F\'{\i}sica
Cl\'{a}ssica 3 graus de liberdade. 

\vspace{.4cm}
{\bf Os 3 graus de liberdade dos corpos ``r\'{\i}gido''-indeform\'{a}veis 
de
Born em Relatividade Restrita, s\~{a}o os mesm\'{\i}ssimos 3 graus de
liberdade dos corpos s\'{o}lidos ``em id\^{e}nticas condi\c{c}\~{o}es
f\'{\i}sicas'' da F\'{\i}sica Cl\'{a}ssica}.
     
     \vspace{.4cm}
O problema dificil em Relatividade \'{e} o de descrever o modo de um
corpo passar de um estado para outro. 
Como o modelo de Born do corpo
``r\'{\i}gido''- indeform\'{a}vel o n\~{a}o permite fazer, havia que
procurar um outro modelo, menos restritivo, que permitisse
deforma\c{c}\~{o}es, mesmo no caso limite dos corpos r\'{\i}gidos.
     
     \vspace{.4cm}
Durante bastante tempo tentei encontrar (definir) um modelo
conveniente de corpo r\'{\i}gido que permitisse passar de uns estados
para outros sem sair de um terreno puramente cinem\'{a}tico.

\vspace{.4cm}
A ideia errada que tinha era a de que todos os corpos r\'{\i}gidos
limite teriam cinem\'{a}ticas semelhantes, independentemente da suas
massas. Ensaiei ``gene-
raliza\c{c}\~{o}es'' das transforma\c{c}\~{o}es
de Lorentz em que fazia entrar as acelera\c{c}\~{o}es  at\'{e} que, um
dia, notei que, por raz\~{o}es de an\'{a}lise dimensional, sem a 
introdu\c{c}\~{a}o 
de mais uma constante fundamental n\~{a}o poderia 
obter f\'{o}rmulas consistentes.
      
   \vspace{.4 cm}
Finalmente, em 1968, veio ao de cima uma  forma\c{c}\~{a}o de
engenheiro mec\^{a}-
nico e convenci-me que o problema era um problema de Elasticidade.
   
   \vspace{.4cm}
Notei que, aceitando a defini\c{c}\~{a}o de corpo r\'{\i}gido
relativista que naturalmente se impunha, a de {\it``corpo em que as ondas
de choque se propagam com a velocidade $c$''}, o simples estudo do
choque de uma barra r\'{\i}gida contra uma parede permitia
determinar a sua lei el\'{a}stica.

   \vspace{.4cm}
Feitos os c\'{a}lculos (apresentados no Ap\^{e}ndice),
usando primeiro a conserva\c{c}\~{a}o da quantidade de movimento e
depois a conserva\c{c}\~{a}o da energia, cheguei, pelos dois
caminhos, \`{a} lei seguinte el\'{a}stica dos corpos r\'{\i}gidos:

\vspace{.4cm} 
\begin{equation}
p=\frac{\rho_{0} ^{0}c^{2}}{2}(\frac{1}{s^{2}}-1),\ \rho_{0} 
=\frac{\rho_{0} ^{0}}{2}(\frac{1}{s^{2}}+1)
, \ com \  s = \frac{l}{l_{0}} 
\end{equation}
   
\vspace{.4cm}
($\rho_{0}^{0}$ - densidade no referencial pr\'{o}prio do material 
r\'{\i}gido
n\~{a}o deformado; $\rho_{0}$ - idem do material r\'{\i}gido deformado; p 
-
tens\~{a}o interna; $l_{0}$- comprimento pr\'{o}prio da barra n\~{a}o 
deformada;
$l$ - idem da barra deformada).
  
  \vspace{.4 cm}
Esta lei \'{e} muito curiosa pois mostra que um material r\'{\i}gido
se pode esticar ilimitadamente sem nunca se partir.
   
   \vspace{.4 cm}
Nos casos a duas e tr\^{e}s dimens\~{o}es, para os materiais
isotr\'{o}picos com coeficiente de Poisson nulo temos as
generaliza\c{c}\~{o}es:
 
 \vspace{.4cm}   
\begin{equation}    
    p_{xx}  =\frac{\rho_{0} ^{0}c^{2}}{4}(\frac{1}{s_{x}^{2}}-1)
    (\frac{1}{s_{y}^{2}}+1) ,
\ \rho_{0} =\frac{\rho_{0} ^{0}}{4}(\frac{1}{s_{x}^{2}}+1)
(\frac{1}{s_{y}^{2}}+1)         
 \end{equation}

\vspace{.4cm}    
\begin{equation}    
    p_{xx}  =\frac{\rho_{0} ^{0}c^{2}}{8}(\frac{1}{s_{x}^{2}}-1)
    (\frac{1}{s_{y}^{2}}+1)(\frac{1}{s_{z}^{2}}+1) ,
\end{equation}
\[
\rho_{0} =\frac{\rho_{0} ^{0}}{8}(\frac{1}{s_{x}^{2}}+1)
(\frac{1}{s_{y}^{2}}+1) (\frac{1}{s_{z}^{2}}+1)         
 \]

(sendo $s_{x}$, $s_{y}$ e $s_{z}$ as deforma\c{c}\~{o}es nas 
direc\c{c}\~{o}es
principais do tensor das deforma\c{c}\~{o}es aqui adoptadas para
direc\c{c}\~{o}es dos eixos de coordenadas).
  
  \vspace{.4 cm}
No caso de um l\'{\i}quido ``rigido'' - com incompressibilidade
m\'{a}xima - as f\'{o}rmulas s\~{a}o:

 \vspace{.4cm}
\begin{equation}
p=\frac{\rho_{0} ^{0}c^{2}}{2}(\frac{1}{s_{v} ^{2}}-1),
\ \rho_{0} =\frac{\rho_{0} ^{0}}{2}(\frac{1}{s_{v} ^{2}}+1)
, \ com \  s_{v} = \frac{v}{v_{0}} 
\end{equation}
   
   \vspace{.4cm}

($v_o$ - volume no referencial pr\'{o}prio do l\'{\i}quido n\~{a}o
deformado; $v$ - idem do l\'{\i}quido deformado).
  
  \vspace{.4 cm}
Usando estes ``materiais'', podemos, em princ\'{\i}pio, fazer nos
quadros da Relatividade Restrita e da Relatividade Generalizada,
n\~{a}o unicamente uma Mec\^{a}nica dos Fluidos, mas tamb\'{e}m uma
Elasticidade Relativista dos corpos r\'{\i}gidos.
  
  \vspace{.4 cm}
Os c\'{a}lculos podem, naturalmente, ser muit\'{\i}ssimo complicados,
mas n\~{a}o temos nenhum problema conceptual.
 
 \vspace{.4 cm}
\hspace{5cm}
x x x
  
  \vspace{.4 cm}
Um problema acess\'{\i}vel \'{e} o do estudo em Relatividade
Restrita dos movimentos estacion\'{a}rios de rota\c{c}\~{a}o de um
disco r\'{\i}gido.
   
   \vspace{.4 cm}
Este problema trata-se em Relatividade praticamente do mesmo modo que
em Elasticidade Cl\'{a}ssica, s\'{o} com a diferen\c{c}a de no
resultado relativista se encontrar para cada disco uma velocidade
angular limite $\omega_{L}$ (facilmente calcul\'{a}vel), acima da 
qual n\~{a}o
pode haver movimentos estacion\'{a}rios.
 
 \vspace{.4 cm}
N\~{a}o h\'{a} qualquer paradoxo porque para o disco atingir esta
velocidade limite  $\omega_{L}$ , que corresponde a uma velocidade $c$ 
do seu rastro,
\'{e} necess\'{a}rio fornecer- lhe uma energia infinita.
 
  \vspace{.4 cm}
 \'{E} preciso ter presente  que no disco relativista, como
num volante real, os movimentos estacion\'{a}rios em que os raios se
mant\^{e}m rectil\'{\i}neos, n\~{a}o s\~{a}o os \'{u}nicos.
 Num volante, mesmo nos movimentos livres em que o volante n\~{a}o
ganha nem perde energia, h\'{a} movimentos em que o rastro avan\c{c}a
e recua em rela\c{c}\~{a}o ao cubo central n\~{a}o se mantendo os
raios rectil\'{\i}neos.

  \vspace{.4 cm} 
Num disco relativista com um dado raio  $R_{L}$, ser\'{a}
poss\'{\i}vel impor dum modo for\c{c}ado numa regi\~{a}o central uma
velocidade angular $\omega$, superior \`{a} velocidade angular limite
$\omega_{L}$, sem que na periferia seja atingida a velocidade $c$. Os
raios ir-se-\~{a}o simplesmente enrolando em espiral enquanto o disco
acumula energia.  Terei neste estudo beneficiado de me ter formado
numa escola de engenharia em que ainda se calculavam volantes.
  
  \vspace{.4 cm} 
Pouco depois de ter encontrado as leis el\'{a}sticas relativistas dos
corpos r\'{\i}gidos vi, num artigo de 1968 [ 4 ], uma refer\^{e}ncia
a dois artigos de Mc Crea de 1952 sobre o assunto.  Quando os li, vi
que Mc Crea j\'{a} tinha encontrado a lei el\'{a}stica a uma
dimens\~{a}o dos materiais r\'{\i}gidos por um processo seme-
lhante ao
meu, com a diferen\c{c}a de primeiro ter considerado a
conserva\c{c}\~{a}o da energia [ 5 ], e depois, j\'{a} com um 
colaborador, a
conserva\c{c}\~{a}o de quantidade de movimento [ 6 ].
 
 \vspace{.4 cm}
Publiquei ent\~{a}o  e uma nota nos Compte Rendus [ 7 ] sobre o 
problema do disco e depois uma lettera no Nuovo Cimento [ 8 ], 
em que citei os
artigos de Mc Crea, e escrevi-lhe a perguntar se tinha desenvolvido
os trabalhos de 1952.
Respondeu-me a dizer que n\~{a}o, mas que, como eu bem suspeitara,
ele sempre tinha continuado a pensar no problema do disco.

 \vspace{.4 cm}
O essencial dos resultados expostos foram incluidos na tese [ 9 ].
Uma dedu\c{c}\~{a}o did\'{a}tica da lei (4) encontra-se em [ 10 ].

 \vspace{.4 cm}
Em 1973, num semin\'{a}rio em Paris em que expus estes assuntos, um
dos presentes, creio que Costa de Beauregard, recordou que Langevin
lhe tinha  um dia dito que o problema do disco era um problema de
Elasticidade. Tinha inteira raz\~{a}o.

  \vspace{.4 cm}
Embora em muitos livros apare\c{c}a a f\'{o}rmula:

\vspace{.4 cm} 
\begin{equation}
\rho_{0} \  c^{2}= \rho_{0} ^{0} \  c^{2} + p
\end{equation}

 \vspace{.4 cm} 
frequentemente usada na Mec\^{a}nica dos Fluidos Relativista, que
pode ser obtida por combina\c{c}\~{a}o das f\'{o}rmulas (4) e das
f\'{o}rmulas (7), nunca encontrei separadas as f\'{o}rmulas (4), (5), (6) 
e (7). 
  
  \vspace{.4 cm}
 Penso que a esmagadora maioria dos relativistas as desconhece.

\section{Alguns desenvolvimentos}

Usando as equa\c{c}\~{o}es (4) e a conserva\c{c}\~{a}o da quantidade
de movimento podemos chegar \`{a} equa\c{c}\~{a}o invariante numa
transforma\c{c}\~{a}o de Lorentz da vibra\c{c}\~{a}o de uma barra
r\'{\i}gida homog\'{e}nea que, como seria de esperar, \'{e}:

\vspace{.4 cm} 
\begin{equation}
\frac{\partial ^{2} X}{\partial x^{2}} - \frac{1}
{c^{2}}\frac{\partial^{2} X}{\partial t^{2}}= 0.
\end{equation} 

\vspace{.4 cm}
Usando a conserva\c{c}\~{a}o da energia obtemos:. 

\vspace{.4 cm}  
\begin{equation}
{\frac{\partial X}{\partial t}} \left(\frac{\partial^{2}X}
{\partial x^{2}} - \frac{1}
{c^{2}}\frac{\partial^{2} X}{\partial t^{2}}\right)= 0.
\end{equation}

\vspace{.4 cm}
Chegamos a um sistema de duas equa\c{c}\~{o}es em que as
solu\c{c}\~{o}es da primeira s\~{a}o solu\c{c}\~{o}es da segunda. A
quest\~{a}o tem import\^{a}ncia porque desenha o quadro que vai
permitir abordar, como veremos adiante, o problema da transmiss\~{a}o
do calor.
 
 \vspace{.4 cm}
A escrita destas equa\c{c}\~{o}es \'{e} particularmente facil em
Relatividade. 
 
   \vspace{.4 cm}
 Determinados em cada ponto, a partir da
descri\c{c}\~{a}o do movimento, a velocidade $v$ e a deforma\c{c}\~{a}o
$s$ locais, usamos as f\'{o}rmulas (4) para determinar a press\~{a}o  $p$ 
e 
e a densidade  $\rho_{0}$, com o que escrevemos as 
componentes $T_{0}^{ \alpha\beta}$ do tensor 
impuls\~{a}o-energia no referencial pr\'{o}prio local.  
Passamos, em seguida, para as componentes  $T^{ \alpha\beta}$ 
do mesmo tensor no referencial de in\'{e}rcia $S$ em que escolhemos
trabalhar. As duas equa\c{c}\~{o}es $\partial_{\alpha} T^{i\alpha} = 0$
e  $\partial_{\alpha} T^{4\alpha} = 0$  que exprimem, as 
leis de conserva\c{c}\~{a}o,
d\~{a}o-nos as equa\c{c}\~{o}es anteriores
\footnote{ Estas equa\c{c}\~{o}es  aparecem-nos escritas em
coordenadas de Euler que s\~{a}o, assim, as coordenadas prop\'{\i}cias
para o estudo da Elasticidade relativista. A  equa\c{c}\~{a}o
de Alembert  da F\'{i}sica Cl\'{a}ssica, a que habitualmente se chega 
por via da  conserva\c{c}\~{a}o da quantidade de movimento, vem 
escrita em coordenadas de Lagrange. Um exerc\'{\i}cio escolar
aconselh\'{a}vel \'{e} o de procurar chegar a esta equa\c{c}\~{a}o por 
via da
conserva\c{c}\~{a}o da energia.}.

\vspace{.3cm} 
\hspace{5cm}
x  x  x
\newpage
  
Em colabora\c{c}\~{a}o com J. C. Fernandes, demos alguns passos em
frente, como seja a generaliza\c{c}\~{a}o  das f\'{o}rmulas (4)  para o
caso de materiais  el\'{a}sticos mas  n\~{a}o r\'{\i}gidos, 
com a velocidade $v_{l}$ de
propaga\c{c}\~{a}o das ondas de choque longitudinais inferior 
a $c$ [11], e  a escrita da correspondente equa\c{c}\~{a}o do
movimento [ 12 ].
 
 \vspace{.4 cm}
O avan\c{c}o mais importante foi, no entanto, devido a um estudante
que na altura ainda n\~{a}o tinha acabado a licenciatura.
 
 \vspace{.4 cm}
Era interessante estudar a velocidade das ondas transversais nos
materiais do tipo (6). Para isso era necess\'{a}rio escrever as
equa\c{c}\~{o}es do movimento a 3 ou, pelo menos, 2 dimens\~{o}es.

 \vspace{.4 cm}
Conhecia um caminho f\'{\i}sico para o fazer, mas longo e com uma
passagem matem\'{a}tica que n\~{a}o sabia resolver analiticamente
(embora soubesse que era resoluvel pois a sabia resolver por
c\'{a}lculo num\'{e}rico).
Estando a dar uma cadeira de Relatividade do $5^{o}$ ano da
licenciatura em F\'{\i}sica da Faculdade de Ci\^{e}ncias de Lisboa
(1984), referi ocasionalmente o assunto a um estudante, Lu\'{\i}s Bento,
que pouco depois apareceu com o problema resolvido.
  
  \vspace{.4 cm}
Lu\'{\i}s Bento, que seguia paralelamente uma outra cadeira onde aprendia
Mec\^{a}nica Anal\'{\i}tica dos meios cont\'{\i}nuos, teve o
m\'{e}rito de notar que o lagrangeano para estabelecer as
equa\c{c}\~{o}es do movimento de um material com as leis
el\'{a}sticas (6) era, exactamente, a densidade $\rho_{0}$ com o sinal 
menos.
  
  \vspace{.4 cm}
Na posse do bom lagrangeano  ${\cal L} = -\rho_{0}$ , reconstruiu o
trabalho de Herglotz e, usando as t\'{e}cnicas convenientes, 
determinou a velocidade das
ondas transversais, obtendo o resultado:

\begin{equation}
v_{t}= \frac{c}{\sqrt{2}}  \ ou\ seja,\  v_{t}= \frac{v_{l}}{\sqrt{2}}  
\end{equation}

\vspace{.4 cm}
Generalizou, ainda, o problema para o caso de materiais com
coeficiente de Poisson  $\sigma$ diferente de zero  obtendo:

\vspace{.4 cm} 
\begin{equation}
v_{t}=c\ \sqrt{\frac{1-2\sigma}{2-2 \sigma}} \ \  ou 
\  seja \  v_{t}= v_{l} \ \sqrt{\frac{1-2\sigma}{2-2 \sigma}}   
\end{equation}

\vspace{.4 cm} 
resultado que coincide com o da F\'{\i}sica Cl\'{a}ssica.

\vspace{.4 cm}
Este trabalho foi publicado numa revista com projec\c{c}\~{a}o [ 13
], mas n\~{a}o parece ter despertado grande interesse.

\vspace{.4 cm}
O contributo de Lu\'{\i}s Bento, veio, no entanto, corrigir e completar,
73 anos depois, o trabalho de Herglotz. Podemos considerar que,
depois dele, a Elasticidade Relativista \'{e}, no essencial, um
cap\'{\i}tulo completo, com direito a uma presen\c{c}a como qualquer
outro nos Tratados de Relatividade
\footnote{Lu\'{\i}s Bento notou que, na linguagem dos campos, as tr\^{e}s
coordenadas fixas $X^i$ s\~{a}o campos escalares. Com efeito, os seus
valores, que nas situa\c{c}\~{o}es reais podem ser gravados sobre o
material, s\~{a}o independentes dos sistemas de coordenadas.
Independentemente de outro interesse, o trabalho tem o interesse
did\'{a}tico de ilustrar um problema de campos com um objecto t\~{a}o
familiar como  \'{e} para n\'{o}s um corpo s\'{o}lido.}.

\section{Problemas em aberto}

 Os materiais relativistas atr\'{a}s considerados s\~{a}o materiais
isotr\'{o}picos. Nas suas deforma\c{c}\~{o}es as direc\c{c}\~{o}es
principais do tensor dos esfor\c{c}os coincidem com as
direc\c{c}\~{o}es principais do tensor das deforma\c{c}\~{o}es,
propriedade que facilita a escrita das equa\c{c}\~{o}es.
 
 \vspace{.4 cm}
Um passo em frente ser\'{a} estudar os materiais anisotr\'{o}picos e
as respectivas equa\c{c}\~{o}es do movimento, em que a
coincid\^{e}ncia referida j\'{a} n\~{a}o se veri-
fica. Ser\'{a},
possivelmente, tamb\'{e}m interessante  estudar teoricamente todas as
poss\'{\i}veis propriedades de anisotropia dos materiais relativistas
\footnote{ S\~{a}o problemas que n\~{a}o se afiguram faceis, mas em que 
os
dados est\~{a}o todos na nossa frente. O m\'{e}todo aconselh\'{a}vel
ser\'{a} reunir todos os dados na cabe\c{c}a e andar a passear no meio 
deles  \`{a}
espera de que um dia surja a solu\c{c}\~{a}o.}.

 \vspace{.4 cm}
\hspace{5cm}
 x x x 

   \vspace{.4 cm}
Um problema que se relaciona com o da Elasticidade \'{e} o da
transmiss\~{a}o do calor.
 
 \vspace{.4 cm}
\'{E} sabido que a equa\c{c}\~{a}o de Fourier \'{e} uma
equa\c{c}\~{a}o n\~{a}o relativista dado permitir (teoricamente) a
transmiss\~{a}o de sinais e de energia com velocidades superiores a
$c$. V\'{a}rios autores t\^{e}m-se esfor\c{c}ado por encontrar 
acrescentos
e variantes que permitam escrever uma equa\c{c}\~{a}o relativista sem
estes inconvenientes. Todos os que conhe\c{c}o (ver, por exemplo, o
livro relativamente recente de Bressan [14 ] ) continuam - sem
\^{e}xito - a estudar a transmiss\~{a}o do calor num ``meio''
r\'{\i}gido-indeform\'{a}vel.
 
 \vspace{.4 cm}
Ora este ``meio'' \'{e} uma no\c{c}\~{a}o geom\'{e}trica s\'{o}
identific\'{a}vel, como mostra-
mos, com corpos f\'{\i}sicos nas
situa\c{c}\~{o}es particulares de estes se mant\'{e}rem {\it nas mesmas
condi\c{c}\~{o}es f\'{\i}sicas}, que n\~{a}o \'{e} o caso quando
h\'{a} transmiss\~{a}o calor com varia\c{c}\~{o}es no tempo.
   
   \vspace{.4 cm}
O quadro f\'{\i}sico para estudar a transmiss\~{a}o do calor em
Relatividade tem, pois, de ser o da transmiss\~{a}o do calor em meios
deform\'{a}veis, no limite com a lei el\'{a}stica dos corpos
r\'{\i}gidos.
 
 \vspace{.4 cm}
No problema n\~{a}o adiab\'{a}tico da vibra\c{c}\~{a}o de uma barra
com transmiss\~{a}o do calor, para al\'{e}m da fun\c{c}\~{a}o  $X= 
X(x,t)$, 
que descreve a configura\c{c}\~{a}o da barra, temos o problema da
determina\c{c}\~{a}o da temperatura $T=T(x,t)$. Passamos, assim, a ter
um problema com duas inc\'{o}gnitas em vez de uma.
 
 \vspace{.4 cm}
As leis de conserva\c{c}\~{a}o da quantidade de movimento e da
energia condu-
zem-nos nesta nova situa\c{c}\~{a}o, n\~{a}o duas
equa\c{c}\~{o}es como as atr\'{a}s referidas em que as
solu\c{c}\~{o}es de uma s\~{a}o solu\c{c}\~{o}es da outra, mas a duas
equa\c{c}\~{o}es diferentes com duas inc\'{o}gnitas diferentes. 
Tal como no caso adiab\'{a}tico as duas
equa\c{c}\~{o}es aparecem-nos escritas em coordenadas de Euler.
No problema a tr\^{e}s dimens\~{o}es a situa\c{c}\~{a}o \'{e} semelhante
sendo, naturalmente, os c\'{a}lculos bastante mais complicados e
quatro as equa\c{c}\~{o}es.  
  
 \vspace{.4 cm}
O quadro da mec\^{a}nica adiab\'{a}tica a que  
nos referimos transformar-se, assim, num quadro prop\'{\i}cio 
ao estudo  da mec\^{a}nica n\~{a}o adiab\'{a}tica. 
  
  \vspace{.4 cm} 
A t\'{e}cnica para escrever as equa\c{c}\~{o}es \'{e} a mesma 
usada no problema anterior, em que come\c{c}amos por escrever o tensor
impuls\~{a}o-energia nos referenciais pr\'{o}prios locais.
   
   \vspace{.4 cm} 
Mas o que \'{e} o referencial pr\'{o}prio de um meio?
   
   \vspace{.4 cm} 
A defini\c{c}\~{a}o mais corrente, adoptada, por exemplo, por
Lichnerowicz no livro [15], \'{e} a de o referencial pr\'{o}prio de
um meio ser aquele em que as componentes $T^{i4}$ do tensor impuls\~{a}o
energia s\~{a}o nulas.
    
    \vspace{.4 cm} 
Ora esta defini\c{c}\~{a}o \'{e} complemamente inaceit\'{a}vel no
caso de haver transmiss\~{a}o do calor.
    
    \vspace{.4 cm} 
Consideremos o caso de uma barra im\'{o}vel num referencial de in'rcia 
onde h\'{a} um fluxo de calor
$q_{o}$ . 
Qual \'{e} o referencial pr\'{o}prio do meio?  Aquele em que a
barra esta im\'{o}vel e temos $T^{i4} = \frac{q_{0}^{i}}{c} \neq 0 $ , ou 
aquele em que a barra 
esta em movimento e \'{e} satisfeita a condi\c{c}\~{a}o $T^{i4} = 0$?
   
   \vspace{.4 cm} 
A defini\c{c}\~{a}o corrente deixa inteiramente de fora os problemas  
onde h\'{a} 
transmiss\~{a}o do calor. Trata-se de mais um caso em que uma 
defini\c{c}\~{a}o defici-
ente encobriu um cap\'{\i}tulo da F\'{\i}sica.
    
    \vspace{.4 cm} 
Para efectivar a escrita das novas equa\c{c}\~{o}es no caso n\~{a}o 
adiab\'{a}tico \'{e} necess\'{a}rio adoptar  rela\c{c}\~{o}es 
constitutivas que definam convenientemente no quadro relativista 
as propriedades do material, n\~{a}o unicamente as proprie-
dades el\'{a}sticas, 
mas tamb\'{e}m as propriedades termodin\^{a}micas. O problema 
esta em aberto 
\footnote{ Curiosamente, a Relatividade vem chamar a aten\c{c}\~{a}o para
um problema que a F\'{\i}sica Cl\'{a}ssica se permitiu ignorar
durante mais de 150 anos. A equa\c{c}\~{a}o de Alembert da
vibra\c{c}\~{a}o de uma barra el\'{a}stica, de 1750, e a
equa\c{c}\~{a}o de Fourier da transmiss\~{a}o do calor numa barra
indeform\'{a}vel, de 1808, s\~{a}o tradicionalmente  
estudadas em cap\'{\i}tulos
separados.  No entanto, na simples vibra\c{c}\~{a}o de uma barra
met\'{a}lica, verificam-se em simult\^{a}neo os dois fen\'{o}menos
da vibra\c{c}\~{a}o e da transmiss\~{a}o do calor. 
Em todo o rigor,
em F\'{\i}sica Cl\'{a}ssica, como em Relatividade, os dois
fen\'{o}menos t\^{e}m de ser estudados com um sistema conjunto de
equa\c{c}\~{o}es. As rela\c{c}\~{o}es constitutivas necess\'{a}rias
para escrever estas equa\c{c}\~{o}es e as t\'{e}cnicas para as
resolver est\~{a}o muito pouco estudadas em F\'{\i}sica Cl\'{a}ssica.}.

\vspace{.4cm}
    \hspace{5cm}
   x x x 
                  
   \vspace{.4 cm} 
Come\c{c}amos a apreender  a no\c{c}\~{a}o de espa\c{c}o
f\'{\i}sico a partir do sentir do nosso corpo. O espa\c{c}o
que ``compreendemos'' em termos correntes \'{e} aquele em que nos
sentimos a existir, onde nos podemos movimentar, onde encontramos
objectos e fen\'{o}menos, onde podemos imaginar outros objectos a 
existir e extens\~{o}es do nosso corpo.

\vspace{.4cm} 
De todos os objectos que encontramos, os mais familiares e mais
faceis de considerar e manipular s\~{a}o os s\'{o}lidos, 
que reunem as duas qualidades da extens\~{a}o e da  
perman\^{e}ncia.
Um s\'{o}lido a deformar-se continua a ser o
mesmo s\'{o}lido. Assim a Geometria, a Ci\^{e}ncia do espa\c{c}o,
\'{e} na, sua fase inicial, a Ci\^{e}ncia em que se observa e estuda o
comportamento dos s\'{o}lidos.

\vspace{.4cm}
Em F\'{\i}sica Cl\'{a}ssica, 
para apreender a no\c{c}\~{a}o de espa\c{c}o, 
o mais c\'{o}modo \'{e} imaginar corpos
r\'{\i}gido-indeform\'{a}veis supostos  ilimitados 
e a eles associar a no\c{c}\~{a}o de {\it referencial}. 
Permitimo-nos assim considerar in\'{u}meros   referenciais, 
todos com m\'{e}tricas euclideanas, 
com os movimentos relativos mais diversos e a
interpenetrarem-se uns aos outros.

\vspace{.4cm} 
Em Relatividade Restrita o problema  \'{e} mais dificil porque os corpos
s\'{o} se podem manter indeform\'{a}veis com uma m\'{e}trica euclideana
quando im\'{o}veis em referenciais de in\'{e}rcia.  
S\'{o} podemos  imaginar  
corpos s\'{o}lidos com m\'{e}tricas
euclideanas se im\'{o}veis em referenciais de in\'{e}rcia, mas podemos  
considerer outros corpos, com os movimentos  mais diversos, 
desde que dotados de leis el\'{a}sticas (nos modelos mais simples). 
As suas m\'{e}tricas pr\'{o}prias 
s\~{a}o , em geral, n\~{a}o euclideanas e vari\'{a}veis. 

 \vspace{.4cm}
Como estes corpos s\~{a}o mais dificeis de estudar
(sobretudo quando h\'{a} falta de conhecimentos de Elasticidade 
relativista),
nos livros de Relatividade Restrita os s\'{o}lidos aparecem habitualmente 
 
no in\'{\i}cio, sob a forma de barras ou comboios para desempenhar um 
papel a  definir  os referenciais de in\'{e}rcia e, depois, ausentam-se.
 
 \vspace{.4cm}
Em Relatividade Generalizada, os corpos indeform\'{a}veis de m\'{e}trica
plana est\~{a}o de todo  excluidos. 
Consideramos  habitualmente nos modelos riemanianos de espa\c{c}o-tempo 
campos, part\'{\i}culas pontuais e, \`{a}s vezes, fluidos, 
mas quem n\~{a}o conhe\c{c}a Elasticidade relativista est\'{a} 
impedido de neles considerar a exist\^{e}ncia de  s\'{o}lidos. 

  \vspace{.4cm}
 Torna-se  dificil "compreender" a Relatividade Generalizada
quando nela n\~{a}o somos capazes de  considerar a exist\^{e}ncia 
de s\'{o}lidos,
que s\~{a}o, exactamente, os corpos que 
pelas suas qualidades de extens\~{a}o 
e perman\^{e}ncia mais facilmente podemos "sentir" como 
pr\'{o}ximos  do nosso corpo. 
Que, melhor nos  permitem 
"sentir-nos inseridos", e como tal "compreender", o  modelo 
matem\'{a}tico   do espa\c{c}o-tempo em que existimos e duramos
\footnote{ Nos livros de F\'{\i}sica \'{e} frequente dizer-se que 
o observador se encontra num 
determinado ponto geom\'{e}trico definido pelas suas coordenadas. 
O nosso corpo  \'{e} , no entanto, extenso.  \'{E} a partir dele que 
come\c{c}amos a apreender e a elaborar a no\c{c}\~{a} de espa\c{c}. 
Podemo-nos imaginar a aumentar,  ou a diminuir, a entrar num 
formigueiro, ou a olhar o Sistema solar "de fora",
continuando-nos a servir  da   habitual no\c{c}\~{a}o de espa‡o. 
Mas n\~{a}o 
nos podemos imaginar a 
entrar no n\'{u}cleo de um \'{a}tomo, nem a ver "de fora" todo o 
Universo.  
N\~{a}o \'{e}, pois,   surpreendente  que  encontremos 
dificuldades quando procuramos abordar os problemas da microf\'{i}ca, ou 
da grande escala do Universo, com as  no\c{c}\~{o}es de espa\c{c}o 
e de tempo 
come\c{c}adas a elabora a partir do nosso corpo.}.  

  \vspace{.4cm} 
Um espa\c{c}o-tempo dado simplesmente pela sua m\'{e}trica,
sem nele sabermos considerar a exist\^{e}ncia de algo, de uma coisa
real, n\~{a}o \'{e} mais do que uma variedade matem\'{a}tica. Por isso
Einstein falava de um  "molusco" 
para com a sua suposta exist\^{e}ncia atribuir significado f\'{\i}sico 
aos
sistemas de coordenadas.

  \vspace{.4cm}

A proposta deste texto \'{e} a de que, para desempenhar o papel deste 
"molusco", se usem corpos com  
leis el\'{a}sticas  r\'{\i}gidas, ou menos r\'{\i}gidas.  
As equa\c{c}\~{o}es
do seu movimento num espa\c{c}o-tempo remaniano previamente
determinado  s\~{a}o   as equa\c{c}\~{o}es que obtemos a partir de  
${\nabla_{\alpha} T^{\alpha\beta} =0}$.
Quando queremos  ter em conta a influ\^{e}ncia do "molusco" 
sobre a m\'{e}trica do espa\c{c}o-tempo, 
basta-nos considerar adicionalmente as equa\c{c}\~{o}es de Einstein.

 \vspace{.4cm}  
\section{Ap\^{e}ndice}

Consideremos uma barra homog\'{e}nea 
de comprimento $l_{0}$ , densidade no refe-
rencial pr\'{o}prio quando n\~{a}o deformada $\rho_{o}^{o}$,

sec\c{c}\~{a}o  $\Delta S$, que se desloca num referencial $S$ 
com a velocidade $v$.

\vspace{.4cm} 
No referencial $S$ o seu comprimento \'{e}: $l=l_{0}\sqrt{1-\beta^{2}}$

\vspace{.4cm} 
Admitamos que no instante $t_0 = 0$ a barra choca com uma parede
indeform\'{a}vel, ou com uma barra igual vinda do outro lado.

\vspace{.4cm} 
Devemos admitir, em Relatividade como em F\'{\i}sica Cl\'{a}ssica,
que a extre-
midade da frente da barra p\'{a}ra, e que uma onda de choque
com a velocidade $v_{l}$, em sentido contr\'{a}rio a $v$, se propaga ao
longo dela separando uma zona parada e comprimida da zona n\~{a}o
comprimida ainda em movimento.

\vspace{.4cm} 
No caso das barras r\'{\i}gidas no sentido r\'{\i}gido-limite
atr\'{a}s exposto, devemos admitir que $v_{l} = c$.  A onda de choque
atinge, neste caso, a segunda extremidade da barra no instante:

\vspace{.4cm}
\hspace{3cm}
$t_{1}= 
\frac{l_{0}\sqrt{1-\beta^{2}}}{c+v}=\frac{l_{0}}{c}\sqrt{\frac{1-\beta}{1+\beta}}$
\vspace{.4cm}

No instante $t_{1}$ a barra esta toda parada em $S$ e tem o comprimento:

\vspace{.4cm}  
\hspace{3cm}
$l_{c}=\frac{c\  
l_{0}\sqrt{1-\beta^{2}}}{c+v}=l_{o}\sqrt{\frac{1-\beta}{1+\beta}}$ 
\vspace{.4cm}  

Representamos a deforma\c{c}\~{a}o por:

\vspace{.4cm}  
\hspace{4cm}
$s=\frac{l_{c}}{l_{o}} =\sqrt{\frac{1-\beta}{1+\beta}}$      

\vspace{.4cm}  
A partir de $t_{1}$, uma outra onda de choque propaga-se ao longo da 
barra
em sentido contr\'{a}rio, separando uma parte n\~{a}o comprimida e de
novo em movimento, agora com a velocidade  $-v$ , da parte parada e
comprimida. Quando esta nova onda atinge a extremidade da frente da
barra, em $t_{2} = 2t_{1} $, toda a barra esta de novo em movimento com a
velocidade $-v$.

\vspace{.4cm}  
Durante o intervalo $[ 0, t_{2}]$ a barra apoia-se na parede (ou na barra
sim\'{e}-
trica vinda do outro lado). Sendo  $p$ a press\~{a}o que a barra
exerce sobre a parede, a for\c{c}a que a parede exerce sobre a barra
\'{e} $f = - p\Delta S$. Num intervalo $\Delta t$ contido em 
$[ 0, t_{2}]$ a varia\c{c}\~{a}o da quantidade de movimento 
da barra \'{e} dada por:

\vspace{.4cm}
\hspace{3cm}
$\Delta P = - \frac{\Delta S\   l_{0}\  \rho_{0}^{0}\  v}
{t_{1}\sqrt{1-\beta^{2}}}\Delta t = $
$ - \frac{\Delta S \  \rho_{0}^{0} \ c^{2} \beta}{1 - \beta} \Delta t$

\vspace{.4cm}  
Usando a f\'{o}rmula:

\vspace{.4cm}
\hspace{5cm}
$\frac{d P}{d t}= f $

\vspace{.4cm}
v\'{a}lida em Relatividade como em F\'{\i}sica Cl\'{a}ssica obtemos:

\vspace{.4cm}
\hspace{3cm}
$ p=\frac{\rho_{0}^{0} c^{2} \beta}{1 - \beta} = \frac {\rho_{0}^{0} 
c^{2}}{2}(\frac{1}{s^{2}} - 1)$

\vspace{.4cm}  
Este resultado \'{e} v\'{a}lido no caso das compress\~{o}es e das
trac\c{c}\~{o}es, havendo neste caso que imaginar, em vez da parede
que trava a frente da barra, um dispositivo mais sofisticado que
agarre a sua parte de tr\'{a}s.

\vspace{.4cm}  
Interessa ainda conhecermos a densidade $\rho_{0}$ da barra quando
comprimida.

\vspace{.4cm}  
A energia da barra n\~{a}o comprimida e em movimento \'{e} igual
\`{a} energia da barra parada e comprimida. Podemos em 
consequ\^{e}ncia
escrever:

\vspace{.4cm}
\hspace{3cm}
$\frac{\rho_{0}^{0} \ \Delta S\  l_{0}}{\sqrt{1 - \beta^{2}}}= \rho_{0} 
\ \Delta S\  l_{0} \sqrt{\frac{1-\beta}{1+\beta}}$

\vspace{.4cm}
donde tiramos:

\vspace{.4cm} 
\hspace{3cm}
$\rho_{0} = \frac{\rho_{0}^{0}}{1 - \beta}= 
\frac{\rho_{0}^{0}}{2}(\frac{1}{s^{2}}+1)$

\vspace{.4cm}
\hspace{5cm}
              x x x

\vspace{.4cm}  
Podemos chegar \`{a} f\'{o}rmula da press\~{a}o por um outro caminho.

\vspace{.4cm}  
Imaginando uma compress\~{a}o lenta da barra parada num referencial $S$, 
do
comprimento inicial $l_{0}$  at\'{e} ao comprimento final $l_{c}$, e 
igualando o
trabalho fornecido \`{a} varia\c{c}\~{a}o de energia, temos :

\vspace{.4cm} 
\hspace{2cm}
$-\Delta S \int_{l_{0}}^{l_{c}}\ p\ dl = c^{2}\ \Delta S \ l_{0}
\  \rho_{0}^{0}\left( \frac{1}{\sqrt{1-\beta^{2}}} -1\right)$

\vspace{.4cm}  
Derivando em rela\c{c}\~{a}o a $l_{c}$ obtemos:

\vspace{.4cm}
$ p = - c^{2}\ \rho_{0}^{0}\  l_{0} \ 
\frac{d}{d\beta}\ 
\frac{1}{\sqrt{1-\beta^{2}}}\ \frac{d\beta}{d l_{0}} = 
- \frac{c^{2}\ \rho_{0}^{0}\ \l_{0} \beta}{( 1 - \beta^{2})^{3/2}}=
\frac{(1-\beta)^{1/2}(1+\beta)^{3/2}}{-l_{0}}= 
\frac{ c^{2}\ \rho_{0}^{0}\ \beta}{1-\beta}$

  \vspace{.4cm}
com o que confirmamos a f\'{o}rmula anterior.

  \vspace{.4cm}
 \hspace{5cm}
 x  x  x

\vspace{.4cm} 
Um outro processo para determinar $p$ ' o de encarar o 
problema do ponto de vista do referencial inicial da barra. 
Neste referencial a parede, que avan\c{c}a sobre a barra com 
a velocidade $-v$, exerce sobre ela a press\~{a}o $p$ durante 
o intervalo de tempo $\Delta t'$.

\vspace{.4cm}
\hspace{4cm}
$ \Delta t' =\frac{ 2 l_{0}}{ v+c} =\frac{ 2 l_{0}}{c (1+\beta)}$  

\vspace{.4cm}
No final a velocidade da barra \'{e}:

\vspace{.4cm}
\hspace{4cm}
$ v_{f} =c \beta_{f} = \frac{-2v}{1+ \frac{v^{2}}{c^{2}}}$ 
 
 \vspace{.4cm} 
Igualando a varia\c{c}\~{a}o da energia da barra no intervalo  $\Delta 
t'$ 
ao trabalho recebido e a varia\c{c}\~{a}o da quantidade 
de movimento ao 
impulso temos :

\vspace{.4cm}
\hspace{1.2cm}
$\Delta E= m_{0} c^{2} \left( \frac{1}
{\sqrt{1 -\beta_{f}^{2}}} - 1 \right) = \Delta S\  l_{0}\  \rho_{0}^{0}  
\ c^{2}\  ( \frac{ 2 \  \beta^{2}}{1 -\beta^{2}})=p\  \Delta S\  v\  
\Delta t' $

\vspace{.2cm}

e 

\vspace{.2cm}
\hspace{2cm} 
$\Delta P = \frac{ m_{0} c\ \beta_{f}}
{\sqrt{ 1 - \beta_{f}^{2}}}= -  p\ \Delta S\ \Delta t' $  
 
 \vspace{.4cm}
  Simplificando, em ambos os casos encontramos para $p$ o resultado 
anterior. 
 
\vspace{.6cm}   
 Os  c\'{a}lculos  apresentados  neste Ap\^{e}ndice   
 podem perfeitamente ser feitos como   
 exerc\'{\i}cios nos cursos elementares de Relatividade.

\end{document}